\documentclass[RNAAS]{aastex62}
\usepackage{gensymb}

\begin{document}

\title{The IAC Stripe82 Legacy Survey: improved sky-rectified images}

\correspondingauthor{Javier Rom\'an}
\email{jroman@iac.es}

\author{Javier Rom\'an}
\affiliation{Instituto de Astrof\'isica de Canarias, c/ V\'ia L\'actea s/n, E-38205, La Laguna, Tenerife, Spain}
\affiliation{Departamento de Astrof\'isica, Universidad de La Laguna, E-38205 La Laguna, Tenerife, Spain}

\author{Ignacio Trujillo}
\affiliation{Instituto de Astrof\'isica de Canarias, c/ V\'ia L\'actea s/n, E-38205, La Laguna, Tenerife, Spain}
\affiliation{Departamento de Astrof\'isica, Universidad de La Laguna, E-38205 La Laguna, Tenerife, Spain}



\section{The IAC Stripe82 Legacy Survey}

The combination of better observational strategies and new instrumentation has allowed the building of large imaging surveys one order of magnitude deeper than current popular datasets such as the Sloan Digital Sky Survey (SDSS). Such advances have opened up the possibility of exploring the low surface brightness Universe with unprecedented precision \citep[see e.g.][]{2016ApJ...823..123T} allowing detailed studies of very low surface brightness galaxies, intra-cluster light and galactic Cirri (among other topics). Amid this new set of deep imaging surveys, the IAC Stripe 82 Legacy Survey \citep{2016MNRAS.456.1359F} is playing a significant role \citep[see e.g.][]{2017A&A...597A.134M,2017ApJ...836..191T, 2017MNRAS.468..703R, 2017MNRAS.468.4039R,2017MNRAS.470..427P}.

The IAC Stripe 82 Legacy Survey is a new co-addition of the SDSS Stripe 82 data \citep{2009ApJS..182..543A}, especially reduced to preserve the faintest surface brightness features of this data set. The survey maps a 2.5 degree wide stripe along the Celestial Equator in the Southern Galactic Cap (-50\degree $<$ R.A. $<$ 60 \degree, -1.25 \degree $<$ Dec.$<$ 1.25 \degree) with a total of 275 square degrees in all the five SDSS filters \textit{(u,g,r,i,z)}. The new reduction includes an additional (deeper) band \textit{(rdeep)}, which is a combination of \textit{g, r} and \textit{i} bands. The average seeing of the Stripe82 dataset is around 1 arcsec. The mean surface brightness limits are $\mu_{lim}$[3$\sigma$,10$\times$10 arcsec$^{2}$] = 27.9, 29.1, 28.6, 28.1 and 26.7 mag arcsec$^{-2}$ for the \textit{u, g, r, i} and \textit{z} bands respectively. The significant depth of the data and the emphasis on preserving the characteristics of the background (sky + diffuse light) through a non-aggressive sky subtraction strategy make this dataset suitable for the study of the low surface brightness Universe. In this research note we present a new data release with improved sky-rectified images. This new data-set is published on the survey webpage ({\href{http://www.iac.es/proyecto/stripe82/}{http://www.iac.es/proyecto/stripe82/}) and is publicly available for the community.

\section{The new sky-rectified co-adds}

Although the coadded images of the IAC Stripe 82 Legacy Survey were carefully reduced, they still contain some residuals along the direction of the drift-scan which is along the Right Ascension. These residuals are illustrated in Fig. \ref{fig:Rectified_S82} and are the result of the different sky brightness in  individual exposures that compose the final co-addition. Due to the careful treatment of the sky in the co-adding, the different sky brightness are preserved in the final stacked images showing residuals that can present brightness (in the worst case) as bright as 26 mag arcsec$^{2}$. A correction of these residuals was already presented in the published version of the survey two years ago \citep{2016MNRAS.456.1359F}. Here we present a new version of the sky-rectified images of the IAC Stripe82 Legacy Survey, improving over the previous correction.

The procedure we have followed for producing better sky-rectified images is as follows. 
To create the masks we use our deepest dataset (i.e. the \textit{rdeep} images).
Masks of all the sources are obtained using SExtractor \citep{1996A&AS..117..393B} on the original coadded images. Later, the masks are enlarged using a Gaussian kernel (with a width of 5 pixels) to  include the missing flux beyond the masks provided by SExtractor. Additionally, to account for the diffuse light on the images (as the ones produced by the scattered light of the sources and the Galactic cirri emission) we create an extra mask using SExtractor in background mode, masking the areas above a certain threshold. Once the final mask in the \textit{r-deep} band is created, we apply such mask to  the rest of the bands. With this treatment the masked  images only contain pixels which are mainly pure sky, or in the worst case, regions with a very weak contamination by diffuse light.  The masked images are then used to estimate the global sky value. We associate such a value to each image as a reference to the sky in that image. In addition, the sky is measured along each pixel row of the image so as to follow the symmetry of the sky residuals (i.e. horizontal lines). We evaluate the difference between such sky row value and the global value and we remove/add such difference to every row. These differences, which are measured in the masked images, are applied to the images without masking, resulting in the sky-rectified images shown in Fig. \ref{fig:Rectified_S82}.

\begin{figure*}
  \centering
   \includegraphics[width=\textwidth]{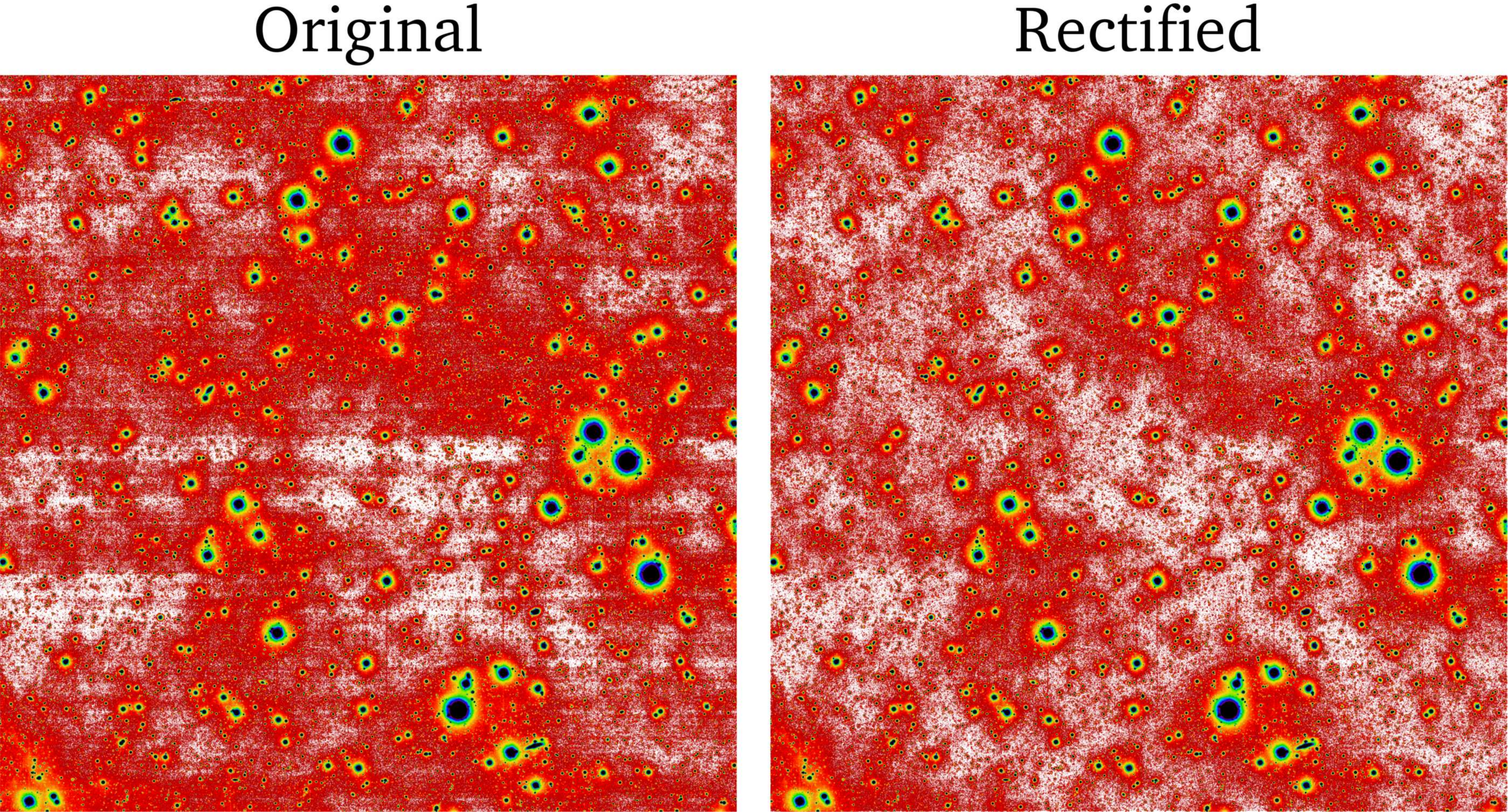}
    \caption{Example of the new sky-rectified images of the IAC Stripe82 Legacy Survey. The images are 0.5\degree$\times$0.5\degree wide. In the left panel the original image  (after sky subtraction) is shown. In the right panel, we show the result of the new sky rectification. Both images are produced using the \textit{rdeep} band (i.e. the combination of \textit{g, r and i} bands).  The images have been convolved with a Gaussian kernel of width of 3 pixels to enhance the contrast. This field, centered at R.A. = 315.25\degree and Dec = 0.50\degree shows a high contamination by Galactic cirri. The cirri are identified with great clarity in the new sky-rectified image.}

   \label{fig:Rectified_S82}
\end{figure*}

The result of the sky rectification is a significant improvement in the quality of the images, showing now a homogeneous sky. The new sky-rectified data allows an accurate photometry of extremely low (fainter than 26 mag arcsec$^{2}$) surface brightness structures, such as the emission by Galactic dust clouds in the Stripe82 area. The sky-rectified images are systematically deeper than the original sky subtracted ones. The improvement in the depth of the images can reach up to 0.1 mag arcsec$^{-2}$ in the case of fields heavily contaminated by diffuse emission.

\acknowledgments

We thank Juergen Fliri for his excellent work on the construction of the IAC Stripe82 Legacy Survey. We also thank Nushkia Chamba for her comments.

\end{document}